\documentclass[pre,floats,aps,superscriptaddress,showpacs]{revtex4}

\usepackage{amsmath}
\usepackage{graphicx}

\begin{document}

\title{ {\bf Thermal convection in mono-disperse and bi-disperse 
granular gases: A simulation study}}

\author{Daniela Paolotti}
\affiliation{Dipartimento di Fisica, Universit\`a di Camerino and
Istituto Nazionale di Fisica della Materia,
Via Madonna delle Carceri, 62032 Camerino, Italy}
\author{Alain Barrat}
\affiliation{Laboratoire de Physique Th{\'e}orique, Unit{\'e}
Mixte de Recherche UMR 8627, B{\^a}timent 210, Universit{\'e} de
Paris-Sud, 91405 Orsay Cedex, France}
\author{Umberto Marini Bettolo Marconi}
\affiliation{Dipartimento di Fisica, Universit\`a di Camerino and
Istituto Nazionale di Fisica della Materia,
Via Madonna delle Carceri, 62032 Camerino, Italy}
\author{Andrea Puglisi}
\affiliation{University ''La Sapienza'',
Physics Department and INFM Center for Statistical Mechanics and
Complexity (SMC), P.le A. Moro 2, 00185 Rome, Italy}


\begin{abstract}
  We present results of a simulation study of inelastic hard-disks vibrated in
  a vertical container.  An Event-Driven Molecular Dynamics method is
  developed for studying the onset of convection. Varying the relevant
  parameters (inelasticity, number of layers at rest, intensity of the
  gravity) we are able to obtain a qualitative agreement of our results with
  recent hydrodynamical predictions. Increasing the inelasticity, a first
  continuous transition from the absence of convection to one convective roll
  is observed, followed by a discontinuous transition to two convective rolls,
  with hysteretic behavior. At fixed inelasticity and increasing gravity, a
  transition from no convection to one roll can be evidenced. If the gravity
  is further increased, the roll is eventually suppressed. Increasing the
  number of monolayers the system eventually localizes mostly at the bottom of
  the box: in this case multiple convective rolls as well as surface waves
  appear. We analyze the density and temperature fields and study the
  existence of symmetry breaking in these fields in the direction
  perpendicular to the injection of energy. We also study a binary mixture of
  grains with different properties (inelasticity or diameters). The effect of
  changing the properties of one of the components is analyzed, together with
  density, temperature and temperature ratio fields.  Finally, the presence of
  a low-fraction of quasi-elastic impurities is shown to determine a sharp
  transition between convective and non-convective steady states.

\end{abstract}
\maketitle


\section{Introduction}

During the last two decades there has been an upsurge of interest for
the physical mechanisms which control the behavior of granular media,
i.e. systems consisting of a large number of macroscopic grains, such as
sand, cereals, powders, etc~\cite{review} These materials play an
important role in many industrial and technological processes and in
natural phenomena and their handling has developed into a
multi-billion-dollar industry. Much experimental and theoretical effort
has been spent on understanding their behavior under a variety of
conditions.  Among the most frequently studied systems are the so called
granular gases~\cite{gases}, obtained by subjecting to an external
driving force an assembly of grains, so that their behavior resembles
that of a molecular gas. However, important differences between the
micro-world and the macro-world, i.e. between the atomic scale and the
millimeter scale, render the analogy incomplete, so that one cannot
infer their properties from the knowledge of the molecular level. This
irreducibility stems chiefly from the presence of non conserving
forces. Striking manifestations of the peculiarity of granular gases are
the non Maxwellian velocity distributions, the shear instability, the
cluster formation to mention just a few.

A typical experiment aimed to probe the behavior of granular gases
consists of a vertical container partially filled with spherical
particles which are accelerated by a vertically vibrating
base~\cite{chicago}.  The competition between the dissipation of
kinetic energy, due to inelastic collisions between the grains, and
the energy provided by the external driving force may lead the system
to exhibit a variety of non-equilibrium statistically steady
states. The phenomenology resulting from such a simple experimental
setup is incredibly rich and by no means trivial.  In addition, by
varying the control parameters, such as the number of particles, the
driving frequency and amplitude, the container aspect-ratio, one can
observe the crossover from one dynamical phase to another.  The
exploration of the resulting phase diagram has been conducted by means
of laboratory experiments and numerical simulations.  Several
properties have also been obtained by granular hydrodynamics and
kinetic theory. However, such calculations are difficult as the
walls and the moving base of the container together with the
gravitational field break the translational invariance of the
system. As a result, one observes density and temperature gradients
which render the theoretical analysis highly non trivial. In addition,
the description of the boundary layers is out of the range of
applicability of hydrodynamics.


In the present work we shall focus on the thermal convection
instability in granular
systems~\cite{ramirez,wildman,kumaran,viot,meerson1,meerson2}, which
consists in the appearance of convection rolls due to the competition
between temperature gradients and gravity. It appears to be different
from standard convection, which is induced by boundaries and excluded
volume effects~\cite{chicago,chicago2}. Thermal convection in granular
media has been first observed in 2D simulations~\cite{ramirez},
confirmed by 3D experiments~\cite{wildman} and analytically
investigated in~\cite{meerson1,meerson2}. Some numerical
investigations can also be found in~\cite{viot,kumaran}.  The
underlying mechanism is analogous to Rayleigh-B\`enard convection in
classical fluids~\cite{chandra}, with the remarkable difference that
in a vibrofluidized granular medium the required temperature gradient
sets in spontaneously as a consequence of the interplay between the
collisional dissipation of energy in the bulk of the granular gas and
of the power injected by the vibrating base.

Hereafter, we shall consider the onset of convective rolls in mono and
bi-disperse two-dimensional vibrated granular gases and its similarity
with the corresponding phenomenon in ordinary fluids. In particular we
shall investigate, by means of event-driven Molecular Dynamics
simulations, a series of predictions provided by the linear stability
analysis of hydrodynamic equations~\cite{meerson2} for pure granular
materials. We shall address the limits of very low and very high
intensity of the gravity. In addition, we shall explore the effects of
polydispersity on thermal convection, studying a binary mixture of
granular gases~\cite{Duparcmeur,Losert,Menon,Wildman,Garzo,Clelland,
MontaneroHCS,Equipart,Puglisi2,Pagnani,Vibrated,Daniela} with
different mechanical properties (in particular different coefficients
of restitution).  We shall characterize the standard hydrodynamic
fields, namely velocity, density and temperature, both in the presence
and in the absence of convective rolls.

In section II we define the model employed to simulate the grains in
the vibrated container, discuss the relevant dimensionless control
parameter and recall the hydrodynamic predictions.  In section III we
investigate the pure system and propose a qualitative comparison with the hydrodynamic
predictions~\cite{meerson1,meerson2} at high and low intensity of the
gravity. In section IV we study the binary mixture case, measuring
velocity, density, and temperature fields. In section V we study the
effect of gradually adding a quasi-elastic component to a
mono-disperse system composed of inelastic disks. We observe a sharp
transition between a convective and a non convective regime.  Finally
in section VI we present a brief discussion and the conclusions.

\section{The model}

Let us consider an assembly of grains constrained to move on a
vertical rectangular domain, representing the container, of dimensions
$L_x \times L_z$, and subjected to a gravitational force acting along
the negative $z$ direction.  The grains are idealized as inelastic
hard disks of diameter $\sigma$ and are fluidized by the
movement of the base, which oscillates with frequency
$\nu=\omega/2\pi$ and amplitude $A$. 
The collisions of the particles with the side and
top walls conserve their kinetic energy 
since the latter are immobile, smooth and
perfectly elastic.  Instead, the collisions between grains are
inelastic, and will be represented by means of non constant coefficients
of restitution, $\alpha_{ij}$, which are functions of the
pre-collisional relative velocity along the direction joining the
centers of a pair, $V_n$, according to the formula:

\begin{displaymath}
\alpha_{ij} (V_n) = \left\{
\begin{array}{lll}
1 - (1 - r_{ij}) \left( \frac{|V_n|}{v_0} \right)^\frac{3}{4} & \mbox{
for } & V_n < v_0 \\
      r_{ij} & \mbox{ for } & V_n > v_0
\end{array}
\right.
\end{displaymath}

\noindent
where $i$ and $j$ indicate the species of the colliding particles,
$r_{ij}$ are constants related to the three types of colliding pairs,
$v_0 = \sqrt{g \sigma}$ and $g$ is the gravitational acceleration. At
large relative velocities the functions $\alpha_{ij}$ assume constant
values $r_{ij}$, representing the coefficients of restitution, but
tend to the elastic value $\alpha_{ij}(0)=1$ for 
collisions occurring with vanishing relative velocity. 
We have checked the insensitivity of our results
on the choice of the value of $v_0$.  The use of velocity dependent
coefficients of restitution has the merit of avoiding anomalous
sequences of collisions among ``collapsed'' particles
\cite{poeschel}.
All the lengths are measured in terms of $\sigma$.

The full dynamics consists of a succession of free streaming
trajectories, which in the presence of the gravitational field
have parabolic shapes, and
inter-particle collisions or wall-particle collisions. It is therefore
very convenient to employ the Event Driven Molecular Dynamics.  We
used the same simulation code of reference~\cite{Daniela}.

The domain was divided into cells and in each cell time-averaged
quantities where computed in order to measure the hydrodynamic
fields.  In such a
coarse-grained description we studied the density, $\rho(x,z)$, 
and velocity field, $\bar{v}(x,z)$.
The local fluctuation of the velocity
defines the granular temperature in the cell ${\cal C}_{x,z}$ centered
around the point of coordinates ($x,z$):

$$T(x,z)=\frac{m}{2}\langle |\mathbf{v} - \langle \mathbf{v}
\rangle_{{\cal C}_{x,z}}|^2 \rangle_{{\cal C}_{x,z}}\ ,  
$$ 
where the symbol
$\langle \cdot \rangle_{{\cal C}_{x,z}}$ is an average performed in the cell
${\cal C}_{x,z}$.
In the measurement procedure we discarded an initial transient and
focused on steady state properties.

From the analysis of the time averaged velocity field we identified the
convection rolls without ambiguity by measuring the circulation, $\Phi$,
of the velocity field~\cite{ramirez}: 

\begin{equation} \label{circ_def}
\Phi=\frac{1}{N_{paths}}\sum_{R}\frac{1}{L(C(R))}\oint_{C(R)}\mathbf{v}\cdot d\mathbf{l}.
\end{equation}

The value of $\Phi$ is obtained, in a steady state, by calculating the sum of
the integrals along $N_{paths}$ different circular paths $C(R)$ of radius $R$
and of length $L(C(R))$ around the center of the box.  The circulation
vanishes in the absence of convection or in the presence of an even number of
convection rolls, and is finite for an odd-number of rolls.

To characterize the dynamical states of the system, it is convenient to
introduce three relevant dimensionless parameters~\cite{meerson1,meerson2}.

The first parameter is the Froude number, which quantifies the relative
importance of the potential energy due to the gravity with respect to the
energy input at the base. In many previous numerical 
studies~\cite{ramirez,kumaran} , the base is
``thermal'': a particle colliding with the base is reinjected with a
velocity taken from a Gaussian
distribution of given temperature $T_{base}$, and thus the Froude number is
given by $F_r=\frac{mgL_z}{T_{base}}$. In hydrodynamical
studies~\cite{meerson1,meerson2}, the temperature at the base, $T_0$, is a
boundary condition and $F_r=\frac{mgL_z}{T_0}$.
In our model, the base is really vibrating, and the energy received on average
by a colliding particle is $m A^2 \omega^2$, so that we can choose as
definition for $F_r$:
\begin{equation} \label{froude2}
F_r = \frac{2gL_z}{A^2\omega^2}=\frac{L_z}{z_{max}},
\end{equation}
\noindent
where $z_{max}=A^2\omega^2/2g$ is the maximum height reached by a projectile
launched vertically with initial velocity $v_0 = A \omega$.  It is worth to
comment that $F_r$ is employed in the present analysis instead of $\Gamma = A
\omega^2/g$, which represents the ratio between the acceleration transmitted
by the vibrating wall and the gravitational acceleration. In the case of a
vibrating base, we note that, in contrast with the less realistic case of a
thermal wall~\cite{ramirez,kumaran} which has been used in previous
simulations, the mass of the particles does not appear in the definition of
the Froude number; this is a relevant point when studying mixtures since
different masses do not lead to different Froude numbers. Moreover, the
presence of a realistic vibrating wall, instead of a stochastic hot wall
(which are not equivalent, see e.g. \cite{sotomansour}), represents a more
severe test of the conjectured convective scenario.  In particular, one can
answer the question whether the movement of the base breaks the structures and
the correlations and renders the convective rolls unstable.

The second dimensionless parameter 
relevant for the problem, the
Knudsen number~\cite{meerson1,meerson2} 
\begin{equation}
K=\frac{2}{\sqrt{\pi}}(\sigma L_z \langle n \rangle)^{-1}=
\frac{2}{\sqrt{\pi} N_{layers}}\ 
\end{equation}
is related to the
mean free path and
$N_{layers}=N\sigma/L$ is the number of filled layers at rest.
Finally, in the case of pure (monodisperse) systems, we consider the dissipative parameter
\begin{equation}
R=8qK^{-2}=\pi (1-r) N_{layers}^2
\end{equation}
where $q=(1-r)/2$ is a measure of the inelasticity of the system
and $r$ is  the coefficient of restitution.
$R$ depends both on the inelasticity and on the collision rate,
since $R \rightarrow 0$ if either $r \rightarrow 1$ or 
$N_{layers} \rightarrow 0$. 

It is useful to recall the hydrodynamic predictions concerning the phase
behavior of the system. In~\cite{meerson1,meerson2} it has presented the
following phase diagram (see figure~\ref{diagram}): at fixed $F_r$ and
$K$, convection rolls appear with increasing inelasticity, i.e. if $R$
overcomes a critical value $R_c$. Such a value, $R_c$, is an increasing
function of the Knudsen number, $K$, which in turn decreases with the
number of particles present in the system. With respect to the Froude
number, instead $R_c$ is a non-monotonic function of $F_r$. As shown in
figure~\ref{diagram}, at low $F_r$ (i.e. at low gravity or strong
shaking) $R_c$ first decreases, i.e. convection is easier to obtain as
the gravity increases.  $R_c$ however reaches a minimum and then {\em
increases} as the gravity is further increased. In the next section we
show the main results of numerical simulations, observing good
{\em qualitative} agreement with predictions from hydrodynamics; however, since
vibrating and thermal walls can yield different behavior for the same
value of the parameters~\cite{kumaran}, no quantitative test will be
made.

\section{Convection in mono-disperse systems}

\begin{figure}[ht]
\centerline{
\includegraphics[clip=true,height=6cm]{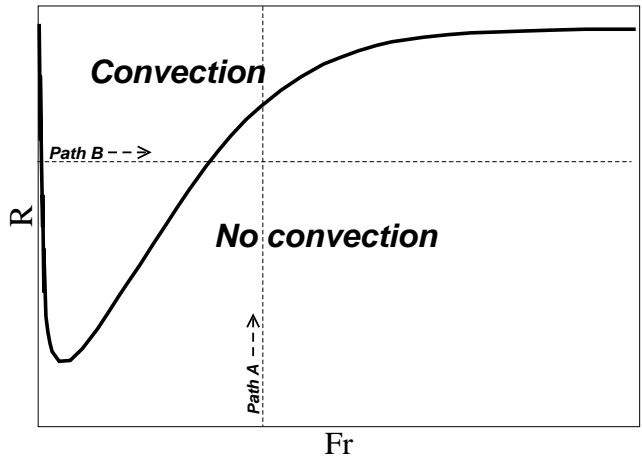}
}
\caption{
Phase diagram in the plane $(F_r,R)$, at fixes $K$, showing the
predictions of hydrodynamic
theories~\cite{meerson1,meerson2}. Convection is expected increasing
$R$, e.g. decreasing the restitution coefficient $r$, as well as
changing $F_r$ in an adequate interval. At too low or too high values
of $F_r$ (e.g. very low or very high gravity) the system does not
reach convection. The two dashed lines indicate the paths followed in
the numerical simulations to verify the predictions. Path A
corresponds to figure~\ref{fig_puro_R}, Path B to
figure~\ref{fig_puro_Fr}.}
\label{diagram}
\end{figure}

\begin{figure}[ht]
\centerline{
\includegraphics[clip=true,height=5cm,angle=-90]{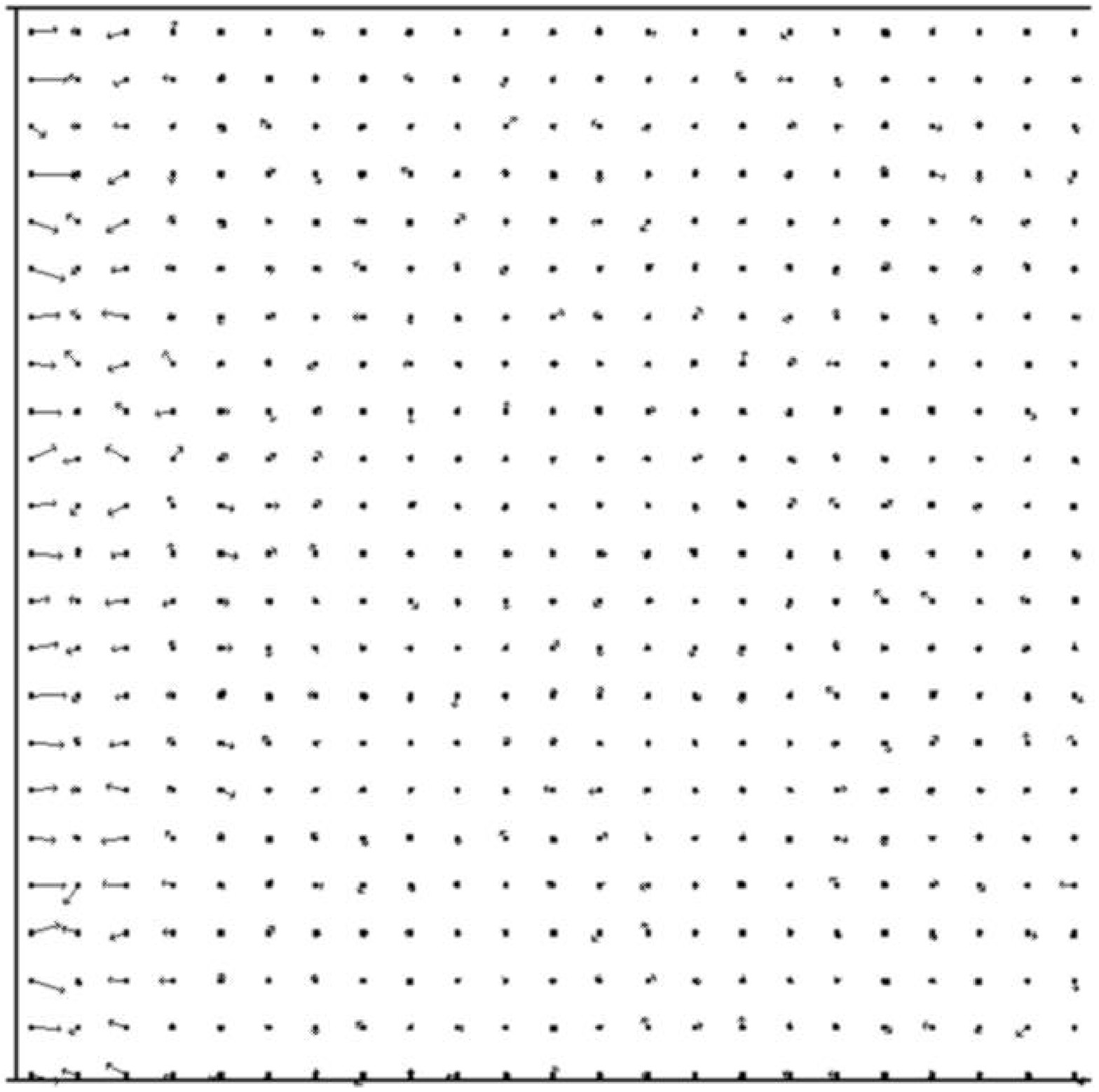}
\includegraphics[clip=true,height=5cm,angle=-90]{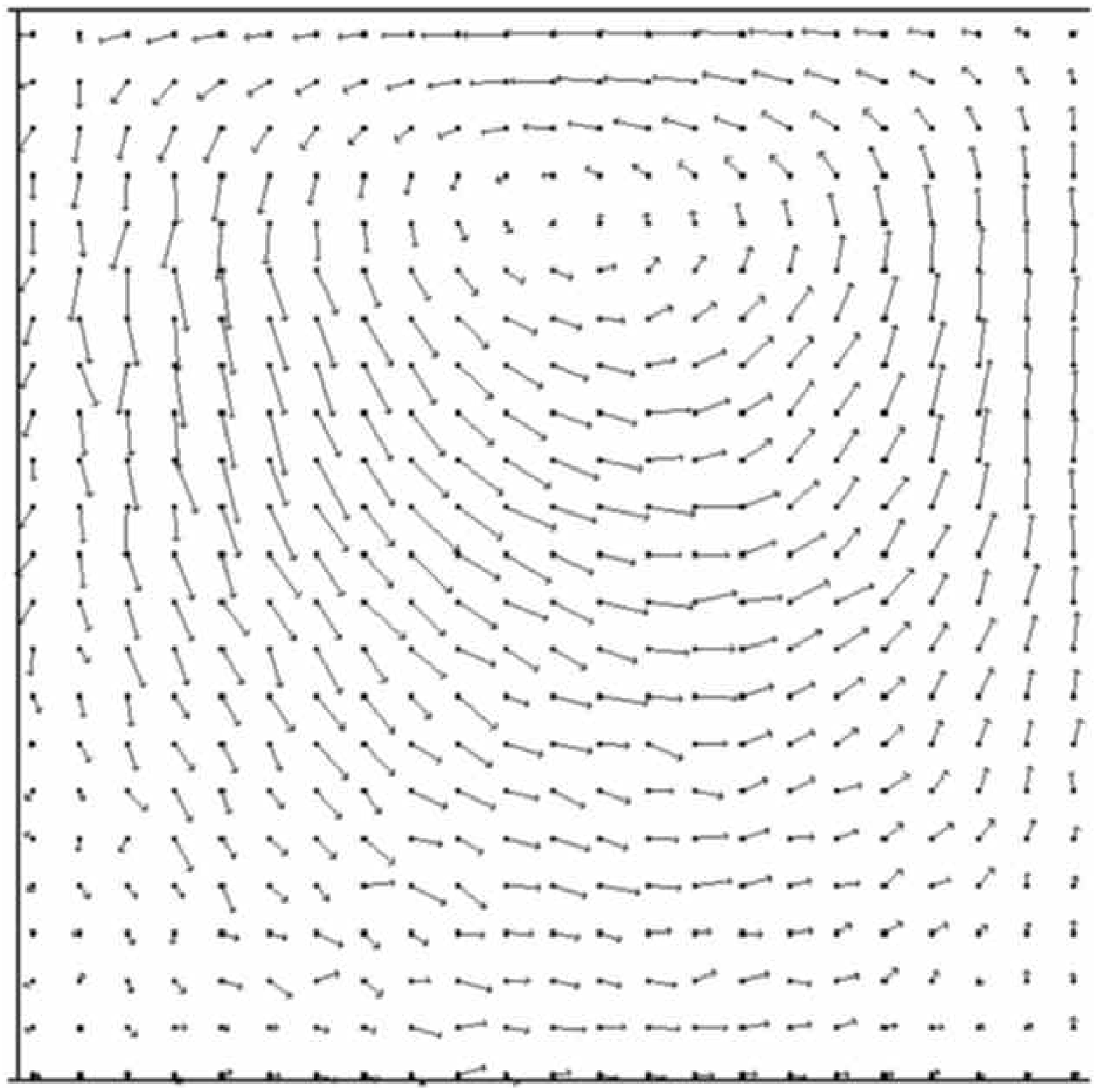}
\includegraphics[clip=true,height=5cm,angle=-90]{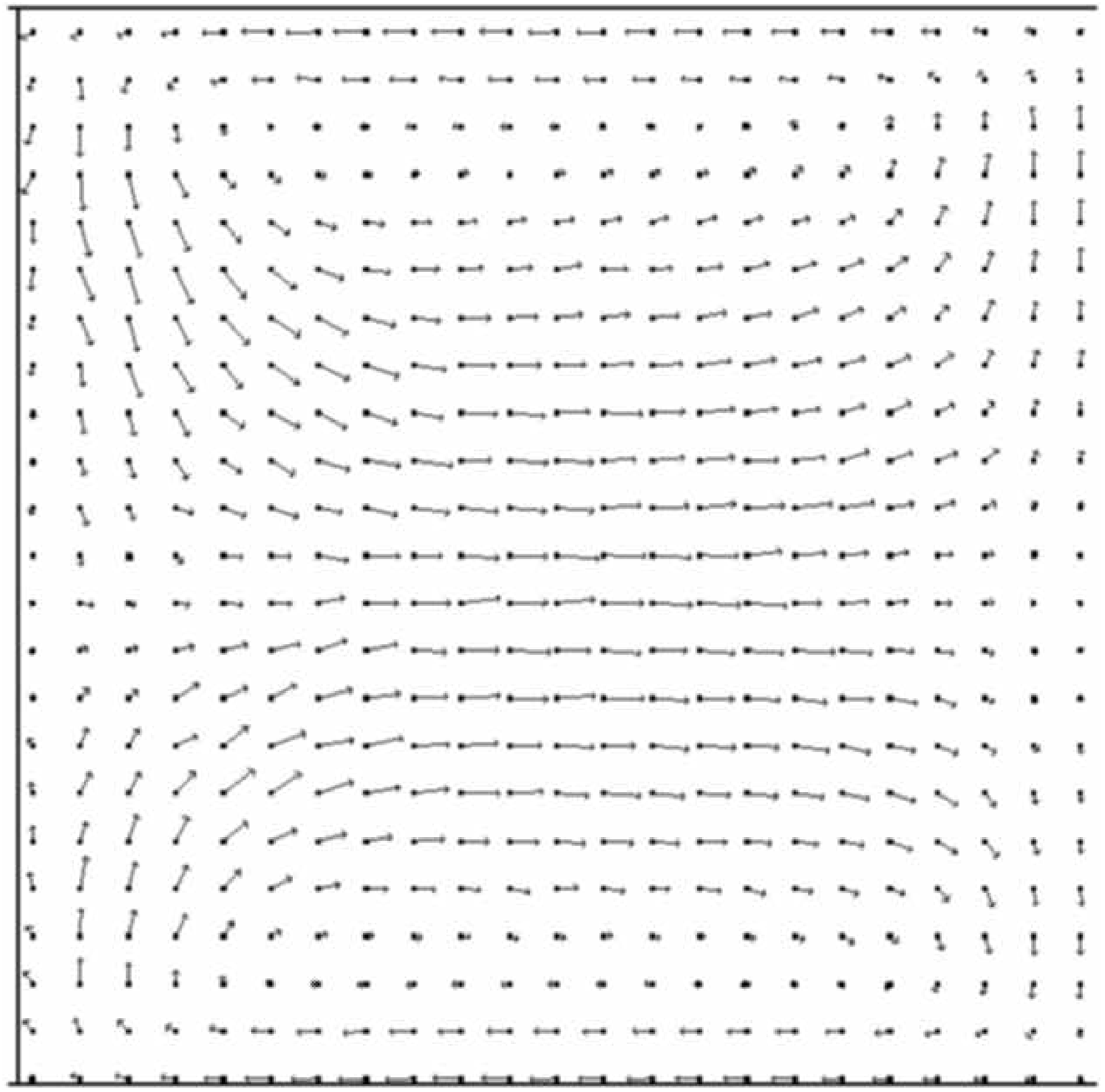}
}
\caption{ {\bf Pure system}: velocity field averaged between $t_1=250$
  and $t_2=300$, in order to show the onset of convection increasing
  $R$, with $F_r=0.36$ and $K=0.071$, $L_x=L_z=113$, $N=1800$, $g=1$,
  $A=0.5$, $\omega=50$. From left to right, the three values of
  restitution coefficient used are $r=0.9992$, $r=0.96$ and $r=0.80$,
  corresponding to $R=0.64$, $R=32$ and $R=159$ respectively.  The
  length of the arrows is normalized so that the maximum velocity
  $v_{max}$ observed is represented by an arrow which has a length
  approximately equal to the distance between two points of the
  lattice. This convention is used in all figures. In these plots the
  value of $v_{max}$ is equal to $14.3$, $14.5$ and $13.6$ (in cm/sec)
  for the case $R=0.64$, $R=32$ and $=159$ respectively.}
\label{fig_puro_R}
\end{figure}

\begin{figure}[ht]
\centerline{
\includegraphics[clip=true,height=10cm,angle=-90]{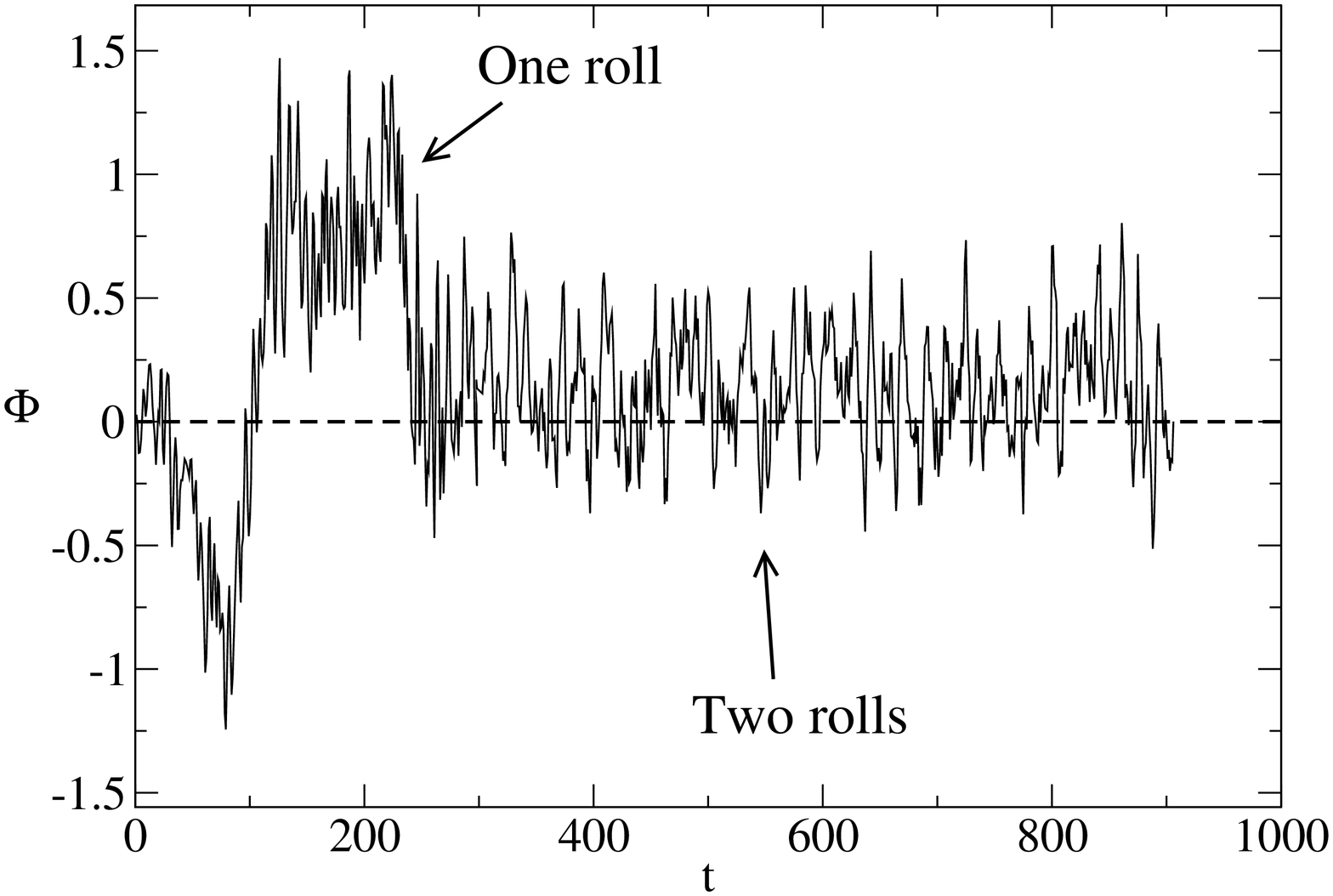}
}
\caption{
{\bf Pure system}: circulation of velocity field $\Phi$ as defined in
the text versus time for the rightmost system of
figure~\ref{fig_puro_R} (i.e. $F_r=0.36$ and $K=0.071$, $L_x=L_z=113$,
$N=1800$, $g=1$, $A=0.5$, $\omega=50$, $r=0.80$, corresponding to
$R=159$). It can be seen that the steady state (identified by $\Phi
\simeq 0$, e.g. two opposite convective rolls) is reached after a long
transient characterized by non-zero circulation (e.g. an odd number of
rolls). In particular the regime between $t=150$ and $t=250$
corresponds to one convection roll very similar to the one depicted in
the central frame of figure~\ref{fig_puro_R}, e.g. for a lower value of $R$.}
\label{fig_puro_R_circ}
\end{figure}

We begin by exploring the phase diagram sketched in
figure~\ref{diagram}, obtained from the
theory~\cite{meerson1,meerson2}, along a plane where $K$ is constant
and moving along two different paths: a line with constant $F_r$ (path
A) and a line with constant $R$ (path B), respectively.

Along path A the effect of increasing $R$ (i.e. of decreasing coefficient of
restitution) is to enhance correlations in the system and render steeper and
steeper the temperature profile in the vertical direction ~\cite{meerson2}.
This leads to the onset of convection and eventually to the increase of the
number of convection rolls. In figure~\ref{fig_puro_R} we display three
pictures corresponding to three different values of $R$. At $R=0.64$ one
observes the absence of convection, whereas a single roll appears at $R=32$.
Finally two rolls are observed for $R=159$. The low values of $F_r$ (low
gravity) and $K$ (low density) allow the particles to explore the whole box,
i.e. regions far from the bottom as well as regions near the base.

Figure~\ref{fig_puro_R_circ}, displays the evolution of the 
velocity field circulation during the transient and the stationary 
regimes in the case of $R=159$: we see that the system first evolves
forming a single convection roll before 
reaching the stationary state characterized by the presence of two
rolls. From fig~\ref{fig_puro_R_circ} we
observe that the formation of a roll occurs sharply.

\begin{figure}[ht]
\centerline {
\includegraphics[clip=true,height=10cm,angle=-90]{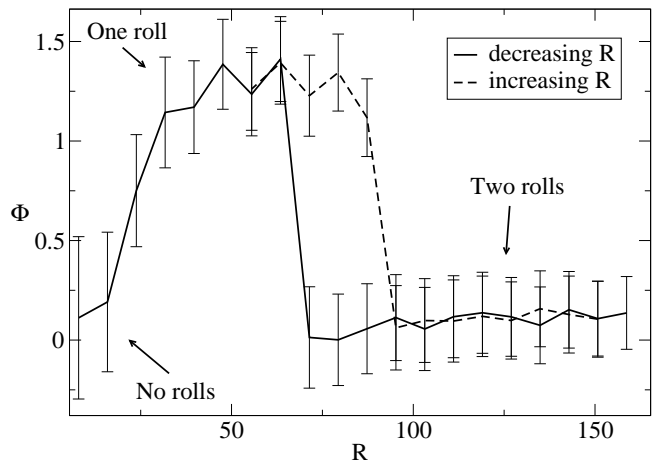}
}
\caption{
{\bf Pure system}: time averages (between time $300$ and $1000$) of
circulation $\Phi$ for different values of restitution coefficient $r$,
i.e. different values of $R$, with $N=1800$, $L_x=L_y=113$, $K=0.071$,
$F_r=0.36$ ($g=1$, $A=0.5$, $\omega=50$). The solid line is obtained
decreasing $R$ and initializing each run with the last configuration of
the previous run. The dashed line is obtained with the same procedure
but increasing $R$. The continuous transition from the absence of
convection to one convective roll, as well as the discontinuous
transition from one roll to two rolls, can be easily recognized. The
second transition shows hysteresis. 
}
\label{fig_puro_order}
\end{figure}

In figure~\ref{fig_puro_order} is shown the circulation (defined in
eq.~\eqref{circ_def}) averaged on times between $300$ and $1000$ and
calculated for different restitution coefficients, between $0.80$ and
$0.99$.  This picture illustrates the presence of two transitions, as
the restitution coefficient is reduced: there appears to be a continuous
transition from absence of convection to one convection roll and a
discontinuous transition from one roll to two rolls. This second
transition shows hysteresis: the solid line is obtained performing
simulations with an increasing restitution coefficient, i.e.  decreasing
$R$, and initializing each run with the last configuration of the
previous run. The dashed line is instead obtained decreasing the
restitution coefficient (increasing $R$). The transition point changes
between the two procedures. As a matter of fact there is a window of
values of $R$ where the system may evolve to one or two rolls, depending
on initial conditions. Such a hysteretic behavior has been pointed out
in~\cite{cordero}, by means of numerical integration of the equations
for granular hydrodynamics.

\begin{figure}[ht]
\centerline {
\includegraphics[clip=true,height=5cm]{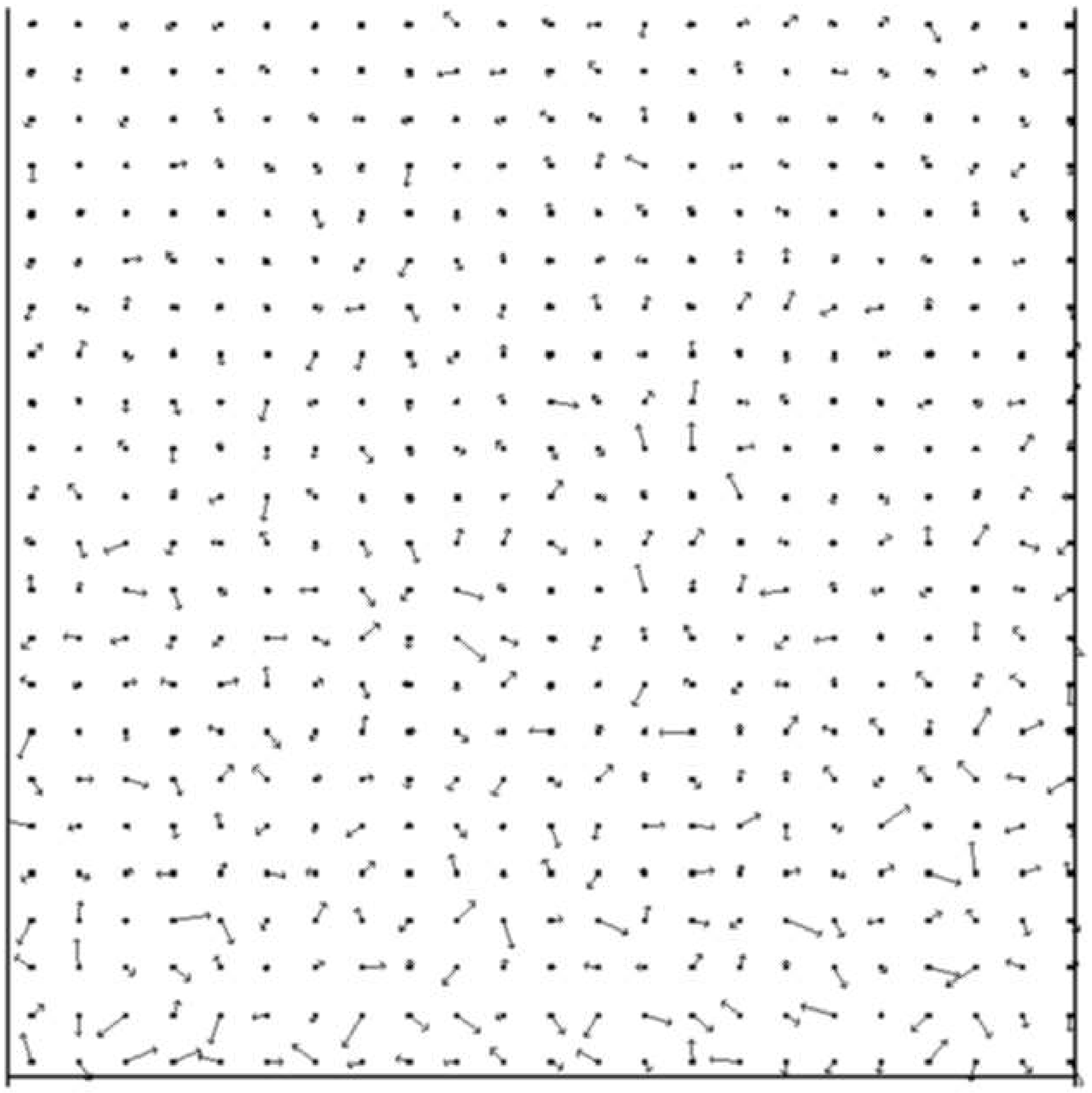}
\includegraphics[clip=true,height=5cm]{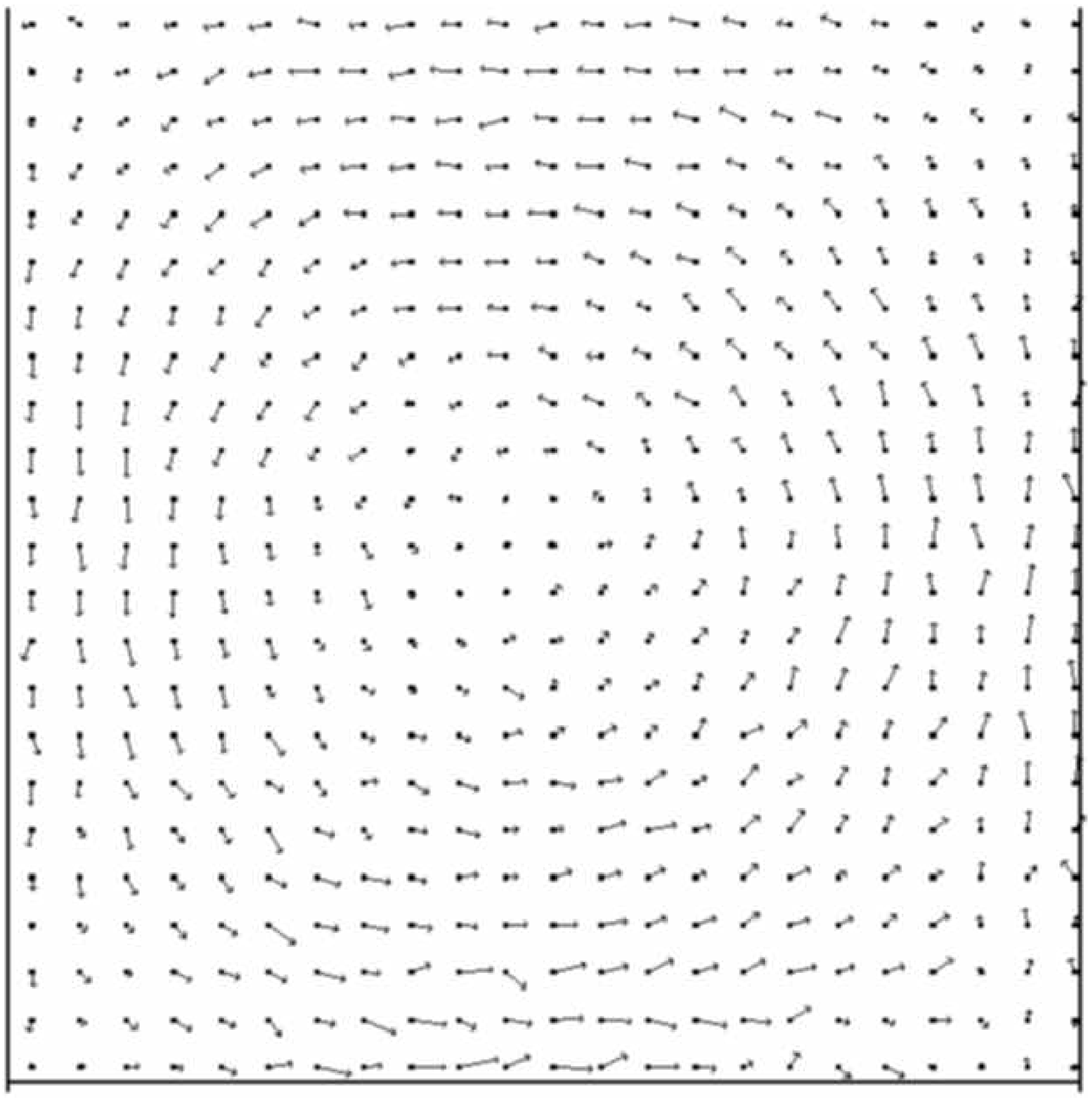}
\includegraphics[clip=true,height=5cm]{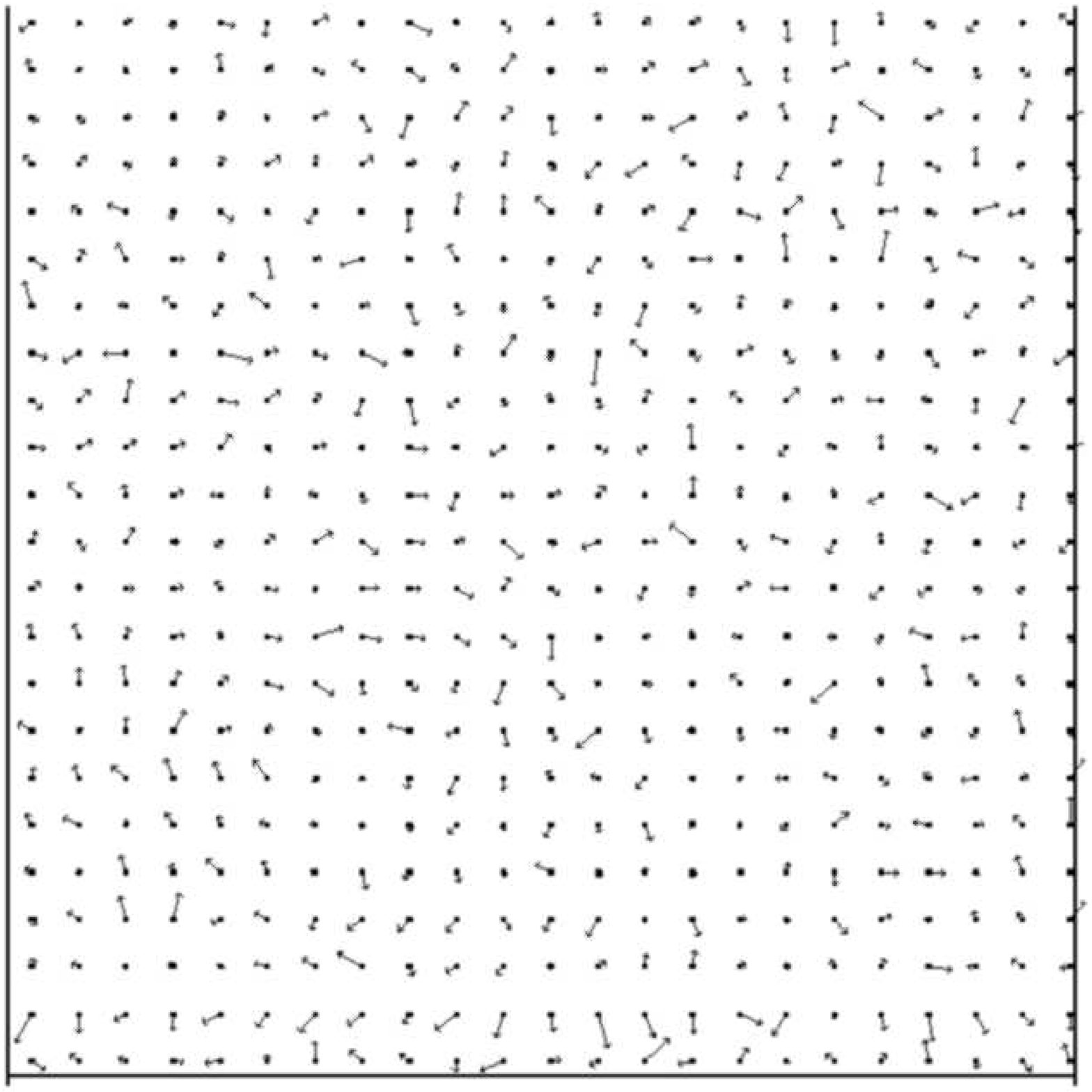}
}
\caption{ {\bf Pure system}: onset and offset of convection (increasing
$F_r$) at $R=3.5$ ($r=0.9955$) and $K=0.07$ fixed, with $L_x=L_z=113$,
$N=1800$, $\sigma=1$, $A=0.5$, $\omega=50$.  The three values of Froude
number are, from left to right, $F_r=1$, $F_r=21$ and $F_r=106$,
corresponding to $g=3$, $g=60$ and $g=300$ respectively. In
correspondence of the lowest value of $F_r$ convection is absent because
$R<R_c$, for $F_r=21$ convection appears because $R>R_c$ and finally for
$F_r=106$ convection disappears again because $R<R_c$. In these figures
the value of $v_{max}$ used to normalize the arrows is $14.3$, $4.5$ and
$2.2$ (in cm/sec) for the case $g=3$, $g=60$ and $g=300$ respectively. }
\label{fig_puro_Fr}
\end{figure}

Along path B ($R$ fixed), instead, the system reveals a non-monotonic
behavior~\cite{meerson1,meerson2}. In fact, the convection stems from
the balance between the vertical negative temperature gradients and
gravity. Therefore, lowering the intensity of gravity (e.g. inclining
the box) reduces such a competition and the value of $R_c$ at which
thermal convection can be observed increases. On the other side, when
$F_r$ (i.e. gravity) exceeds a certain threshold, there is a crossover
to a substantially different regime: the grains tend to remain
localized near the vibrating plate. Such a spatial arrangement
determines an increase in density in the lower part of the box and
tends to inhibit convection, because the packing reduces the mobility
of the grains. For this reason when $F_r$ becomes larger, the value of
the threshold, $R_c$, increases. In figure~\ref{fig_puro_Fr} such a
non-monotonic behavior is shown: one can see three averaged velocity
fields obtained for three different values of $F_r$, while keeping $R$
and $K$ fixed.  To conclude, the patterns obtained by Molecular
Dynamics are consistent with the prediction of~\cite{meerson2} based
on granular hydrodynamics.

\begin{figure}[ht]
\centerline {\includegraphics[clip=true,height=5cm]{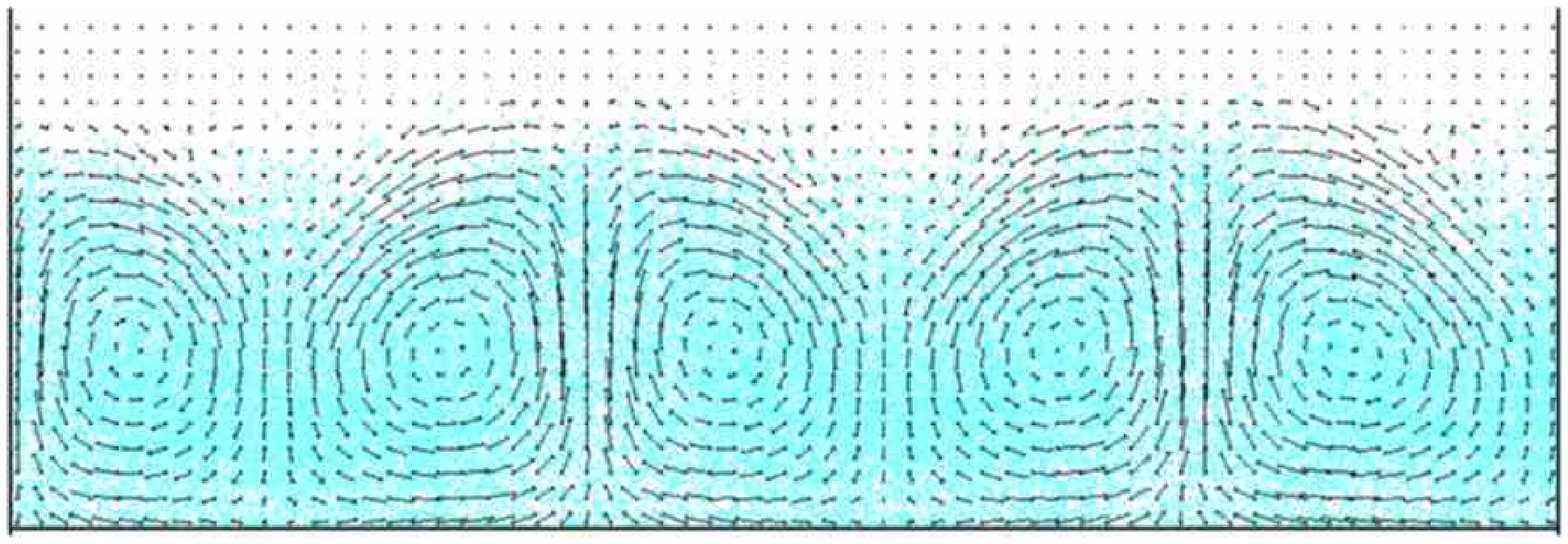}}
\centerline {\includegraphics[clip=true,height=5cm]{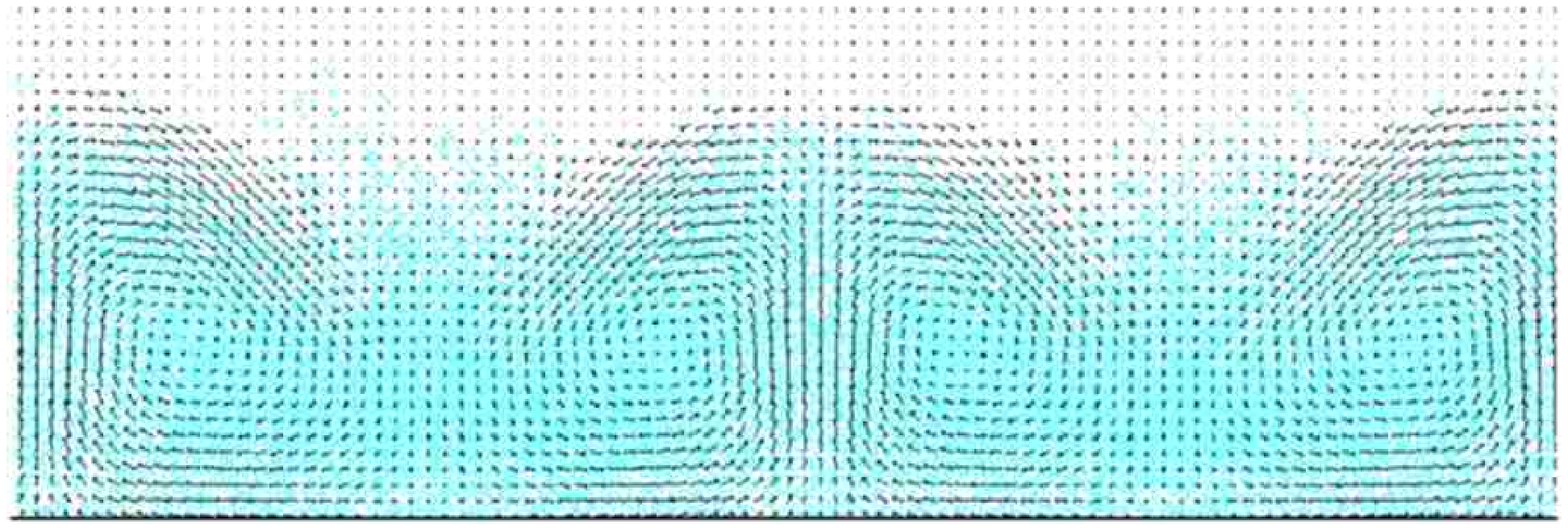}}
\caption{
  {\bf Pure system}: convection at high $F_r$ number (e.g. strong gravity). We
  used $r=0.991$, $L_x=375$ ($L_z$ is irrelevant in this case),
  $N=18000$, $\sigma=1$, $A=0.5$, $\omega=50$, $g=28$, so that $K=0.023$ and
  $R=65$. Top: fixed walls boundary conditions; bottom: periodic boundary
  conditions}
\label{fig_highFr}
\end{figure}

Next, we have tested the prediction of~\cite{meerson2} regarding the
convection at very large values of the Froude number, i.e. when the
particles~\cite{cinesi} are confined in a region adjacent to the basal plate.
In such a case the typical length-scale characterizing the onset of convection
suggests the presence of a multi-roll configuration. In
figure~\ref{fig_highFr} we display the velocity field of a system with large
aspect ratio $L_x/L_z=1.5$ and a very high Froude number, $F_r=22.5$. In this
regime, almost no collisions with the upper wall occur, and $L_z$ is therefore
essentially irrelevant. Remarkably, five convection rolls are present and a
wave-like horizontal profile of the density is observed, revealing a
horizontal density instability (see also~\cite{kumaran}). Note that an odd
number of rolls can only be obtained with fixed walls boundary conditions,
while periodic boundary conditions lead always to an even number of rolls,
because neighboring cells must rotate in opposite directions.

\begin{figure}[ht]
\centerline {
\parbox{1.5cm}{\includegraphics[clip=true,height=4cm]{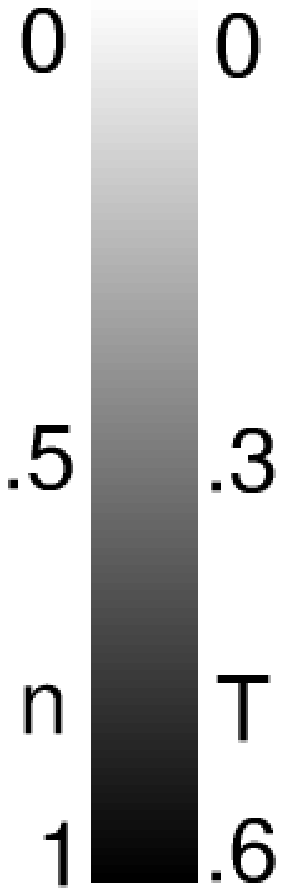}}
\parbox{12cm}{
\includegraphics[clip=true,width=5.5cm]{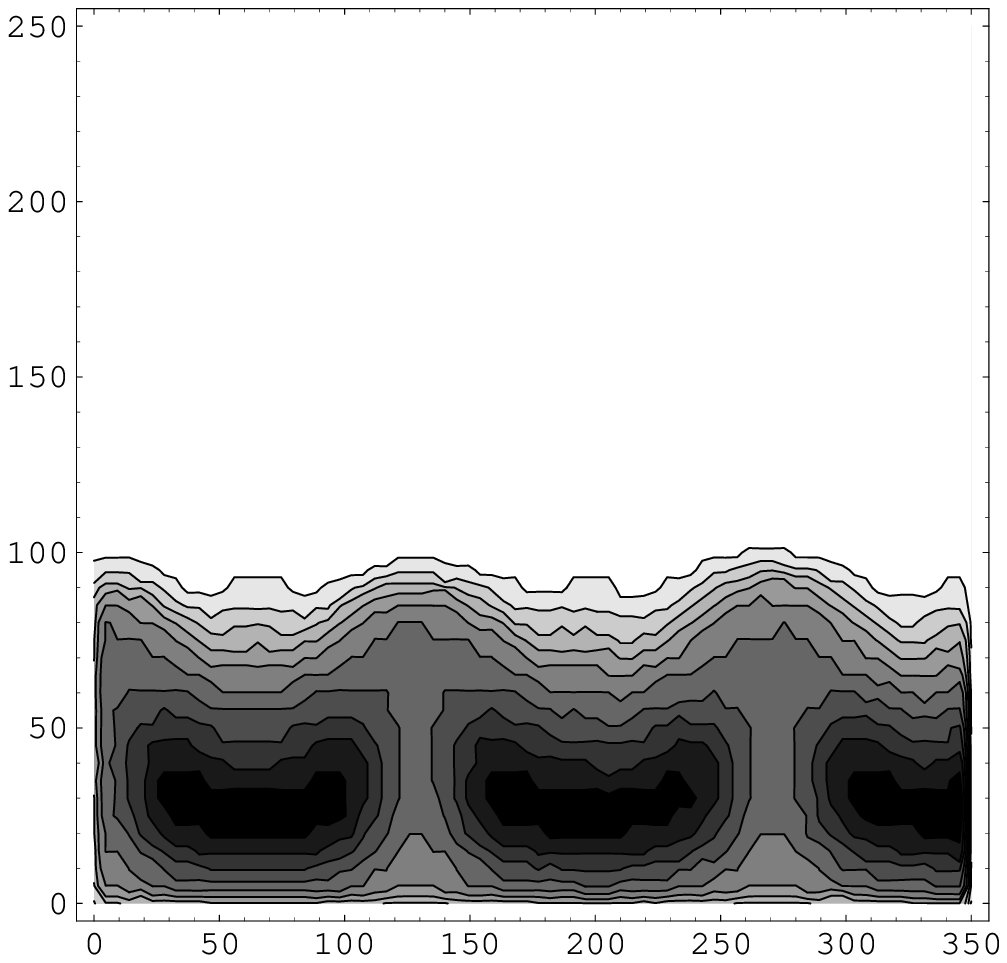} 
\includegraphics[clip=true,width=5.5cm]{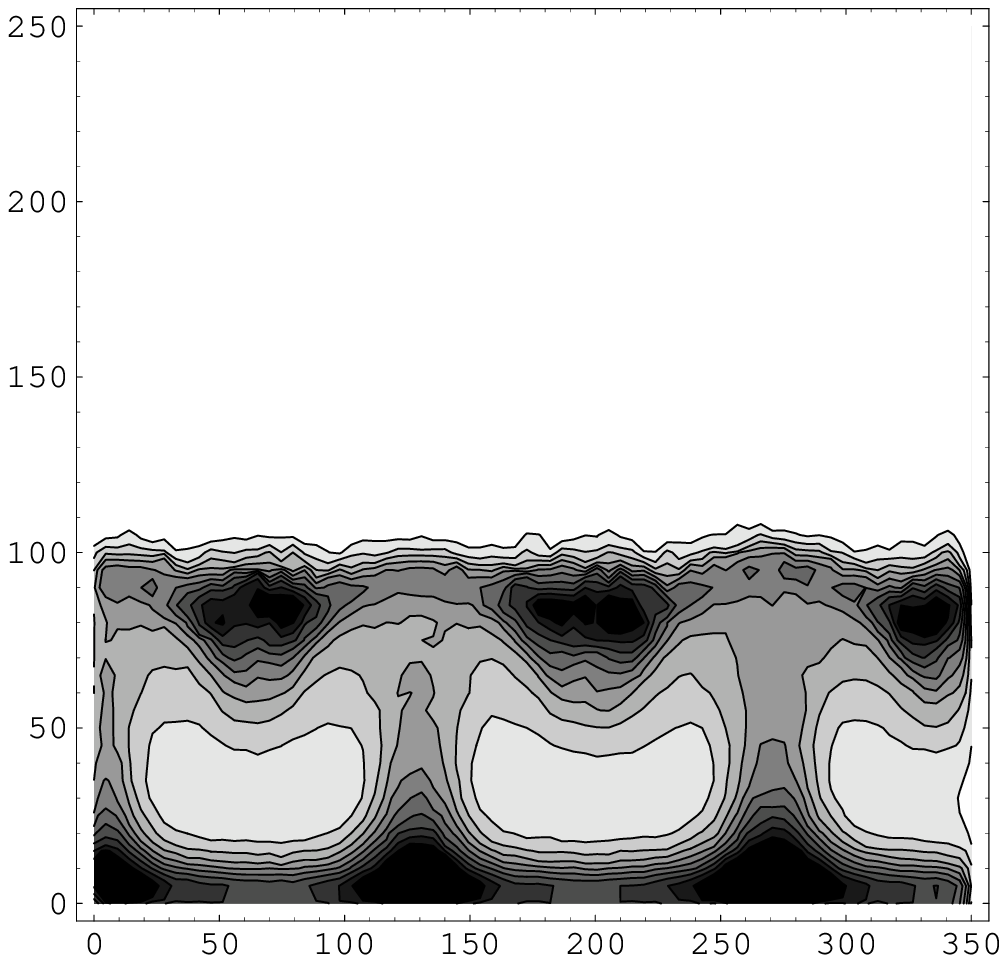}}}
\caption{ {\bf Pure system}: convection at high $F_r$ number (strong
gravity), with the same parameters of figure~\ref{fig_highFr} (with
fixed wall boundary conditions): the density (left) and temperature
(right) fields in arbitrary units.  }
\label{fig_highFr_fields}
\end{figure}

In order to complete our description of the hydrodynamic properties of the
system, we have analyzed the granular temperature and density fields shown in
figure~\ref{fig_highFr_fields} (in the case of fixed lateral walls). Both
fields show clearly the horizontal inhomogeneity: the density instability and
the corresponding temperature instability can be explained in terms of the
peculiar clustering mechanism characterizing granular fluids. Inelastic
collisions, which occur more frequently in dense regions, determine a
reduction of the local temperature and consequently a drop of the local
pressure. The ensuing pressure gradient amplifies the density fluctuation and
more particles flow toward that region.

Note that the three regions, where the density is larger, are located in
correspondence of the three ``valleys'' at the free surface visible in
figure~\ref{fig_highFr}. The temperature field reflects the symmetries of the
density field, and one can see that density and temperature are
anti-correlated: the denser regions are colder. For instance, the three
``islands'' of hot particles correspond to the three lower densities (i.e. the
holes left empty by the depression of the free surface, see again
figure~\ref{fig_highFr}). However, above a certain height (about $100$), the
horizontal instability disappears and both temperature and density decrease
together: this is a region with very low number of particles.

Direct comparison between figures~\ref{fig_highFr} and
\ref{fig_highFr_fields} clearly shows the strong correlation between
the velocity field and the temperature and density fields. In the
region separating two convective rolls, two opposite phenomena can be
observed: (i) if the velocity field is directed upward, i.e. toward
the free surface, the density is relatively low and one observes a
bump on the surface; (ii) if the velocity field is directed downward,
the bottom plate confines the particles and one observes both a
depression in the surface and a high density region close to the
bottom. Each convection roll therefore displays an asymmetric shape
with a high-density region coupled to the downward flow and a
low-density region coupled to the upward flow. We notice that this
phenomenology is slightly different from that observed
in~\cite{kumaran}, where a convective regime and a clustering regimes
where studied separately: our simulations
indicate a strong interplay between density and velocity fields.

\section{Binary mixtures}

In the present section, we shall 
consider how the convective properties of a granular
fluid can be modified by the addition of a second species with
different physical properties.  First, we shall trigger the convection
by tuning the parameters of one of the two species, in the case of
equal concentrations. We also consider the local density ratio, that
accounts for segregation. 

\begin{figure}[ht]
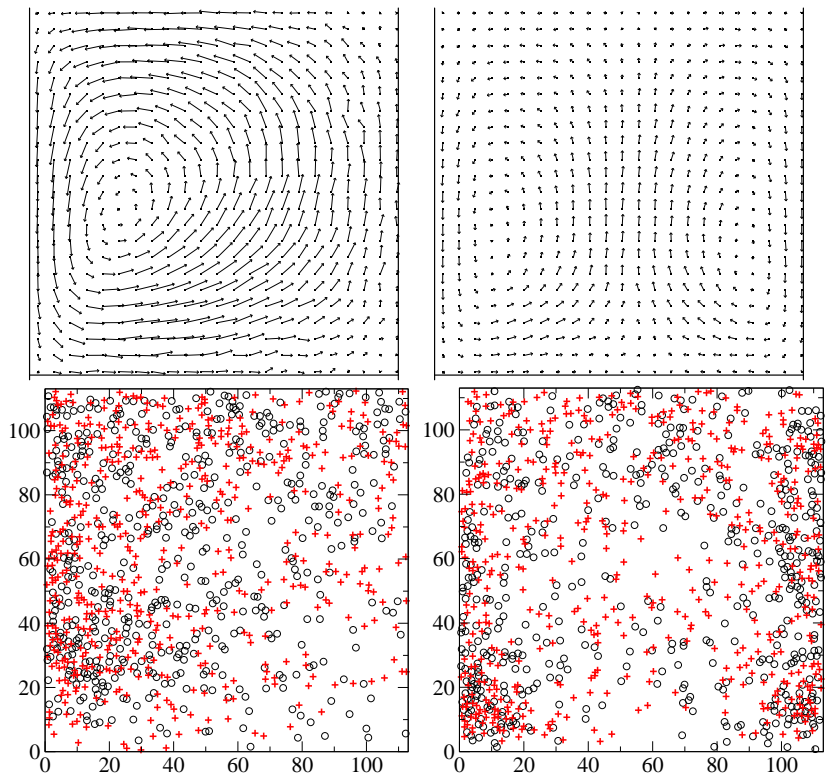

\centerline{\includegraphics[clip=true,width=5cm,angle=-90]{mix_r11fix_r22096_vfield.eps} \hspace{.2cm}
\includegraphics[clip=true,width=5cm,angle=-90]{mix_r11fix_r22060_vfield.eps}}
\centerline{\includegraphics[clip=true,width=5.4cm]{mix_r11fix_r22096_snapshot.eps} 
\includegraphics[clip=true,width=5.4cm]{mix_r11fix_r22060_snapshot.eps}}
\caption{
{\bf Binary mixture}: two different choices of $r_{22}$ determine one or
two convective cells; (above) the velocity field, (bottom) a snapshot
of particles positions (crosses and circles are species $1$ and $2$
respectively). The parameters used are: $\sigma=1$, $L_x=L_z=113$,
$N_1=N_2=550$, $g=1$, $A=0.5$, $\omega=50$, $r_{11}=0.96$, so that
$K=0.12$ and $F_r=0.36$. The two values used for $r_{22}$ are $0.90$,
(left) and $0.60$(right). In the two upper plots the value $v_{max}$ used to
normalize the arrows is $9.5$ and $7.2$ (in cm/sec) for the left and right case respectively. }
\label{fig_r11}
\end{figure}

While non-equipartition can be determined  by differences either of masses,
sizes or inelasticities, we have 
mentioned in the introduction that
the Froude number does not depend on the masses of the particles when
the energy is injected
through a vibrating wall. We shall therefore focus
on mixtures of beads with different inelasticities or sizes.

We shall begin with a mixture consisting of two kinds of grains with
different coefficients of restitution. We denote with $r_{11}$ the
coefficient of restitution relative to collisions among particles both
belonging to species $1$, $r_{22}$ relative to collisions between
particles of species $2$ and $r_{12}$ relative to collisions between
unlike species. For the sake of simplicity, we shall assume
$r_{12}=(r_{11}+r_{22})/2$. Figure~\ref{fig_r11} shows how the
variation of the inelasticity of only one of the two species may
change the convective structure: in particular we reduce $r_{22}$
keeping $r_{11}$ fixed. The species $1$ has a restitution coefficient
$r_{11}=0.96$ equal to the case of figure~\ref{fig_puro_R}, which
displayed a single convection roll. In a first simulation (left frame)
the species $2$ has a restitution coefficient $r_{22}=0.90$, and still
a single convection roll is observed. In a second simulation, with
$r_{22}=0.60$, two convection rolls are found. In the same figure we
also display two corresponding snapshots, in order to give a hint of
the coupling between the rolls and the density: as in the previous
section, regions of high density are coupled to the downward flow of
particles. Moreover the snapshots suggest the absence of segregation,
as expected for dilute systems. The absence of segregation and the
visual inspection of convection rolls suggest a first simple
hypothesis for the behavior of a binary mixture with different
restitution coefficients: an equimolar mixture behaves similarly to a
pure gas with an effective restitution coefficient that is roughly an
average between those of the two components. In fact in the first case
the ``average'' restitution coefficient is still near the value where
a single cell was observed in the pure gas, while in the second case
this effective coefficient (around $0.78$) is similar to the one
corresponding to two rolls in the pure gas (rightmost frame of
figure~\ref{fig_puro_R}).

As far as the temperature ratio field $T_1/T_2$ is concerned, we
cannot present a conclusive picture. In fact, the temperature ratio
profiles that can be efficiently measured in situations where the
systems are horizontally homogeneous ( see for instance
refs. \cite{Menon,Equipart,Daniela}) become extremely noisy due to the presence
of transversal structures and convection. Our results suggest that the
temperature ratio is not appreciably affected by convection rolls or
at least the variation with respect to the case without rolls cannot
be resolved due to the statistical uncertainty.  We recall that in
granular gases the breakdown of equipartition is not necessarily
related to segregation and can be predicted on the basis of the
homogeneous Boltzmann equation for dissipative hard spheres (or
disks), both in the absence and in the presence of external driving
forces.

\begin{figure}[ht]
\centerline {\includegraphics[clip=true,height=5cm]{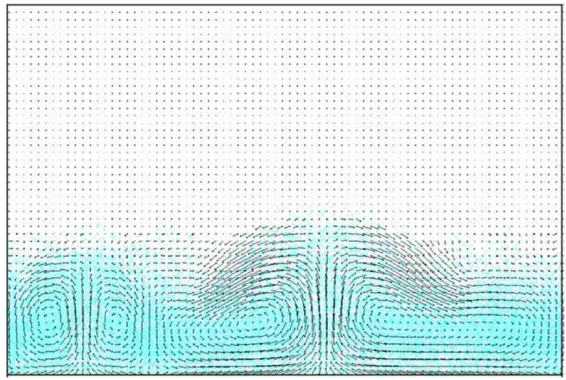}}
\caption{
{\bf Binary mixture}: a mixture of two different sizes
($\sigma_2/\sigma_1=0.8$ not far from $1$) shows differences in the
convective pattern, with respect of figure~\ref{fig_highFr}. The
parameters are the same used for that figure. In this plot the value
$v_{max}$ used to normalize the arrows is $7.3$ cm/sec. }
\label{fig_highFr_mix}
\end{figure}

\begin{figure}[ht]
\centerline {
\parbox{1.5cm}{\includegraphics[clip=true,height=5cm,keepaspectratio]{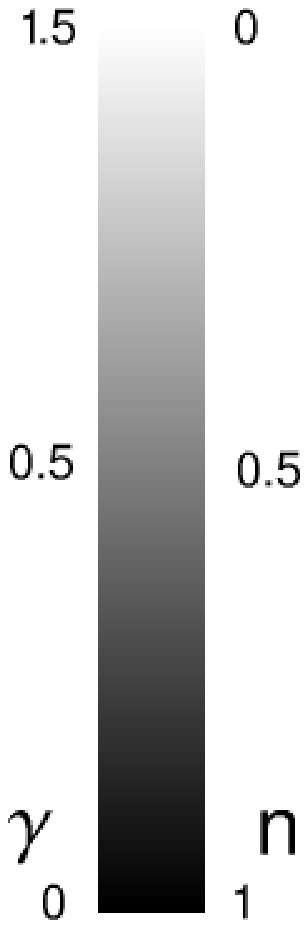}}
\parbox{12cm}{
\includegraphics[clip=true,width=5.5cm,keepaspectratio]{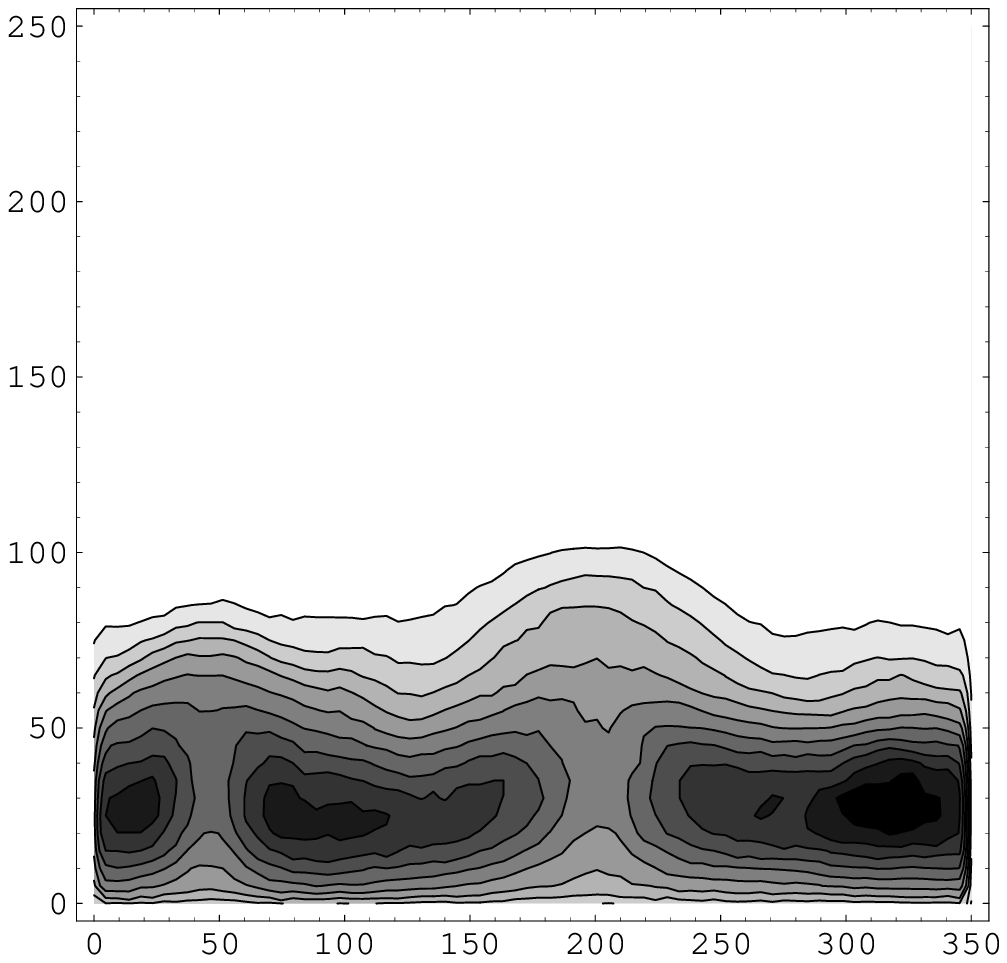}
\includegraphics[clip=true,width=5.5cm,keepaspectratio]{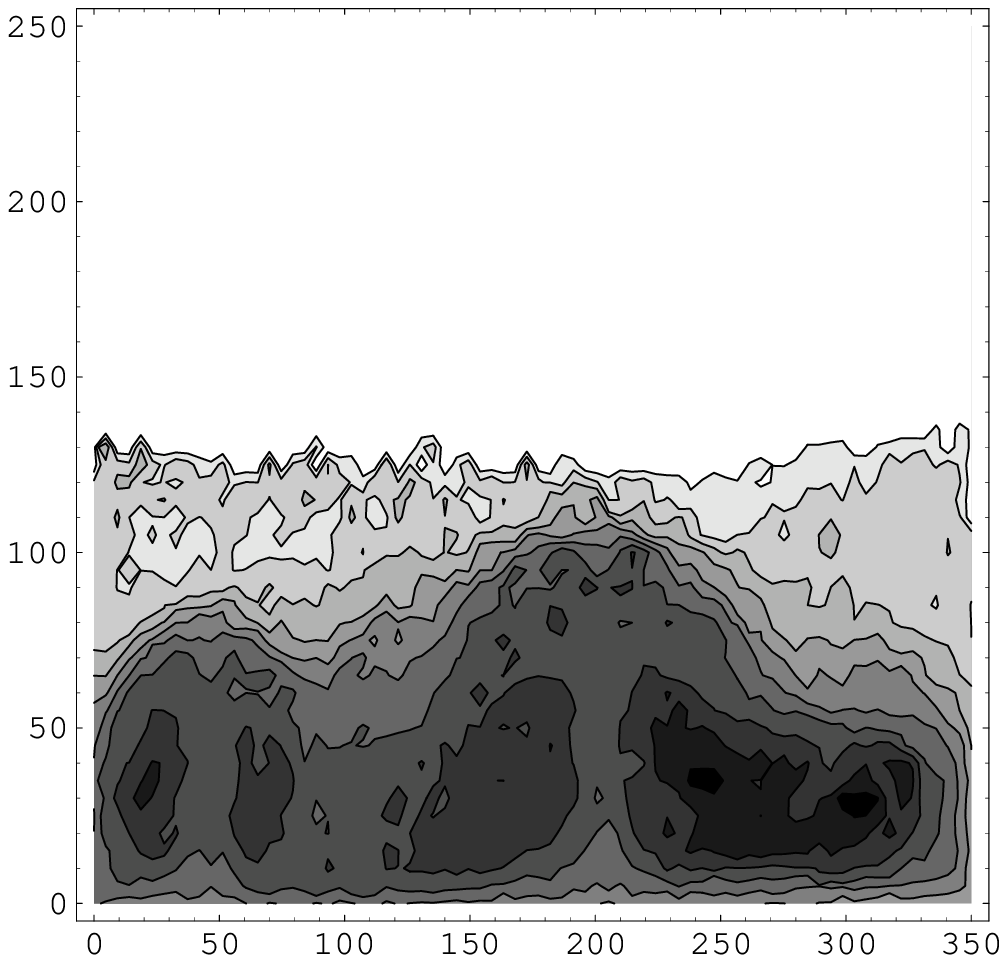}}
}
\caption{
{\bf Binary mixture}: Density fields: left $n_1$ field and right ratio
$\gamma=n_2/n_1$, with the parameters of figure~\ref{fig_highFr_mix}}
\label{fig_highFr_mix_fields}
\end{figure}

Finally, we have studied how to enhance segregation of a mixture, 
in a case
where the global density was higher and the sizes of the 
particles where different.
In particular, we have chosen the parameters corresponding to
figure~\ref{fig_highFr}, but substituted half of the grains with grains
having larger diameter ($\sigma_2/\sigma_1=0.8$). The velocity
field is shown in figure~\ref{fig_highFr_mix}. The field is averaged
over very long times, i.e. few hundreds of oscillations of the plate, 
and the
horizontal asymmetry observed appears to be stationary, although,
in principle, this pattern is expected to be
metastable, in the sense that an infinitely long average should produce
a horizontally periodic field.  Therefore, a slight change in the
size of half of the particles affects the convective field leaving
unaltered the number of rolls, but changing their relative size and
shape.
We have analyzed the density fields of the single species $n_1$ and
$n_2$ and the density ratio $\gamma=n_2/n_1$, obtaining the results in
figure~\ref{fig_highFr_mix_fields}. Segregation is now evident,
appearing as deviations of $\gamma$ from $1$. In particular it can be
noted that at the bottom there are more particles of species $1$ (the
larger particles). This results calls for new experiments in this
dilute ``granular gas'' regime: in fact it is opposite to well known
experimental findings in the densely packed
regime~\cite{chicago}. Moreover, it appears that stronger segregation
is localized in regions of high density.

\section{Non equimolar mixture}

Even in a granular mixture the relative concentration,
$\chi=N_2/(N_1+N_2)$, is expected to play an important role.
Not only it affects the ratio of the partial granular temperatures,
but also can induce a transition from a homogeneous to a convective
velocity regime. This has particular relevance, for instance, in industrial
processes that require the absence or the presence of convection: in the
simulations of this section we show how a small percentage of
``doping'' quasi-elastic particles added to an assembly of more inelastic
particles can totally suppress convection.

\begin{figure}[ht]
\centerline {\includegraphics[clip=true,width=6cm,keepaspectratio]{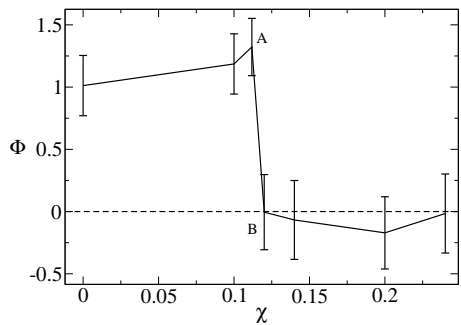}}
\caption{Averaged circulation $\phi$ in a mixture
with $L_x=L_y=113$, $g=1$, $r_{11}=0.96$, $r_{22}=0.9992$,
$N_1+N_2=2200$, corresponding to $K=0.06$, $F_r=0.36$. Doping is
measured as percentage of quasi-elastic particles,
i.e. $N_2/(N_1+N_2)$}
\label{fig_circulation}
\end{figure}

\noindent
By employing as an
order parameter to detect the transition the
circulation $\Phi$, of the velocity field, already defined 
in eq.~(\ref{circ_def}) and varying $\chi$ we obtained the curve
depicted in figure~\ref{fig_circulation}. In these
simulations, we assumed: $r_{11}=0.96$, $r_{22}=0.9992$,
$N=N_1+N_2=2200$, $K=0.06$, $F_r=0.36$. While the pure system is in a
convective regime, a small ``doping'' (the critical value is estimated
to be $\chi_{cr} \approx 0.12$) suppresses convection. This critical
value changes if the relevant parameters of the system (e.g. $K$ or
$F_r$) are changed. For example we have verified that increasing $F_r$ the
value $\chi_{cr}$ increases.

\begin{figure}[ht]
\centerline
{\includegraphics[clip=true,height=5cm,,angle=-90]{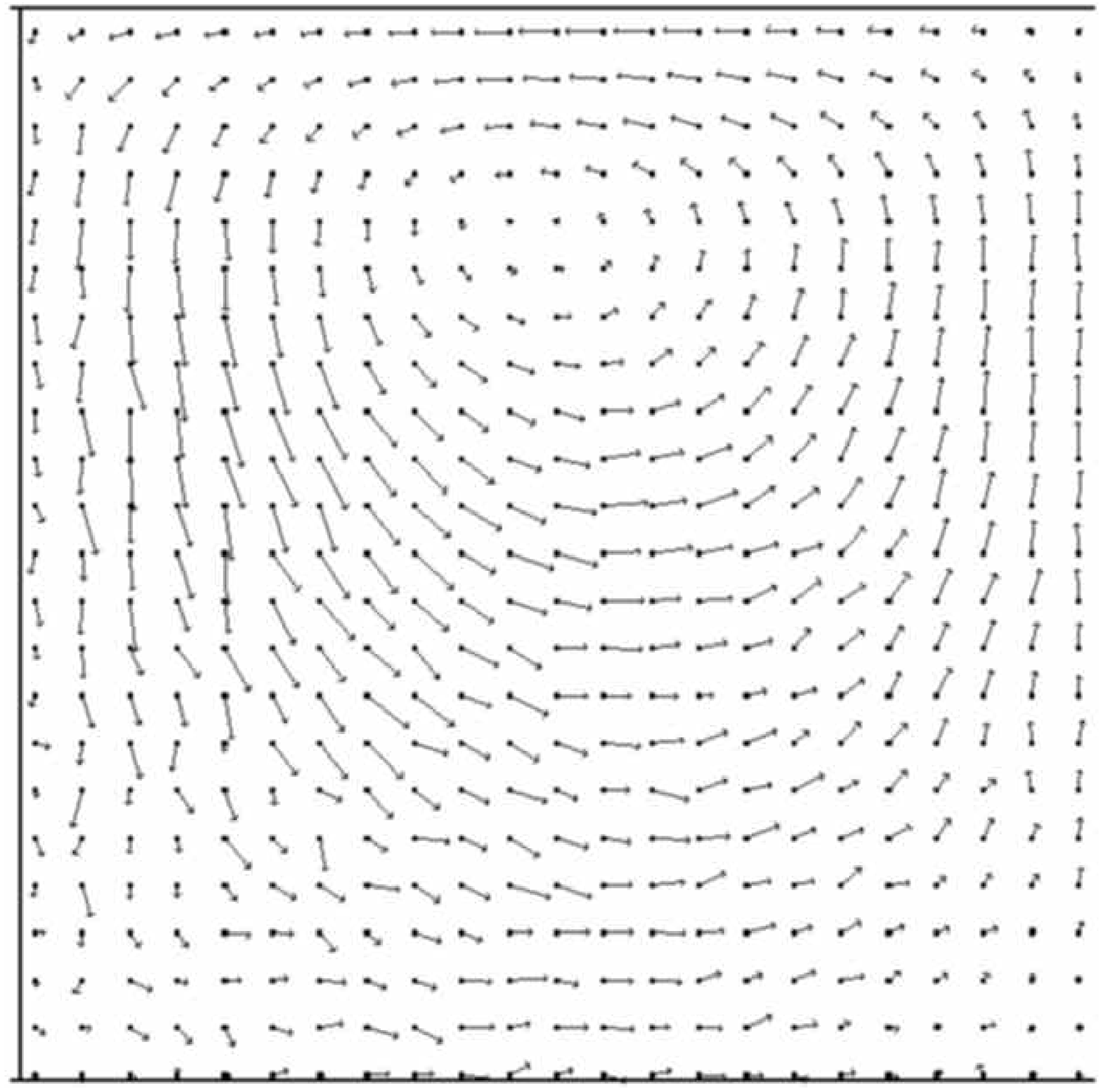}
\includegraphics[clip=true,height=5cm,angle=-90]{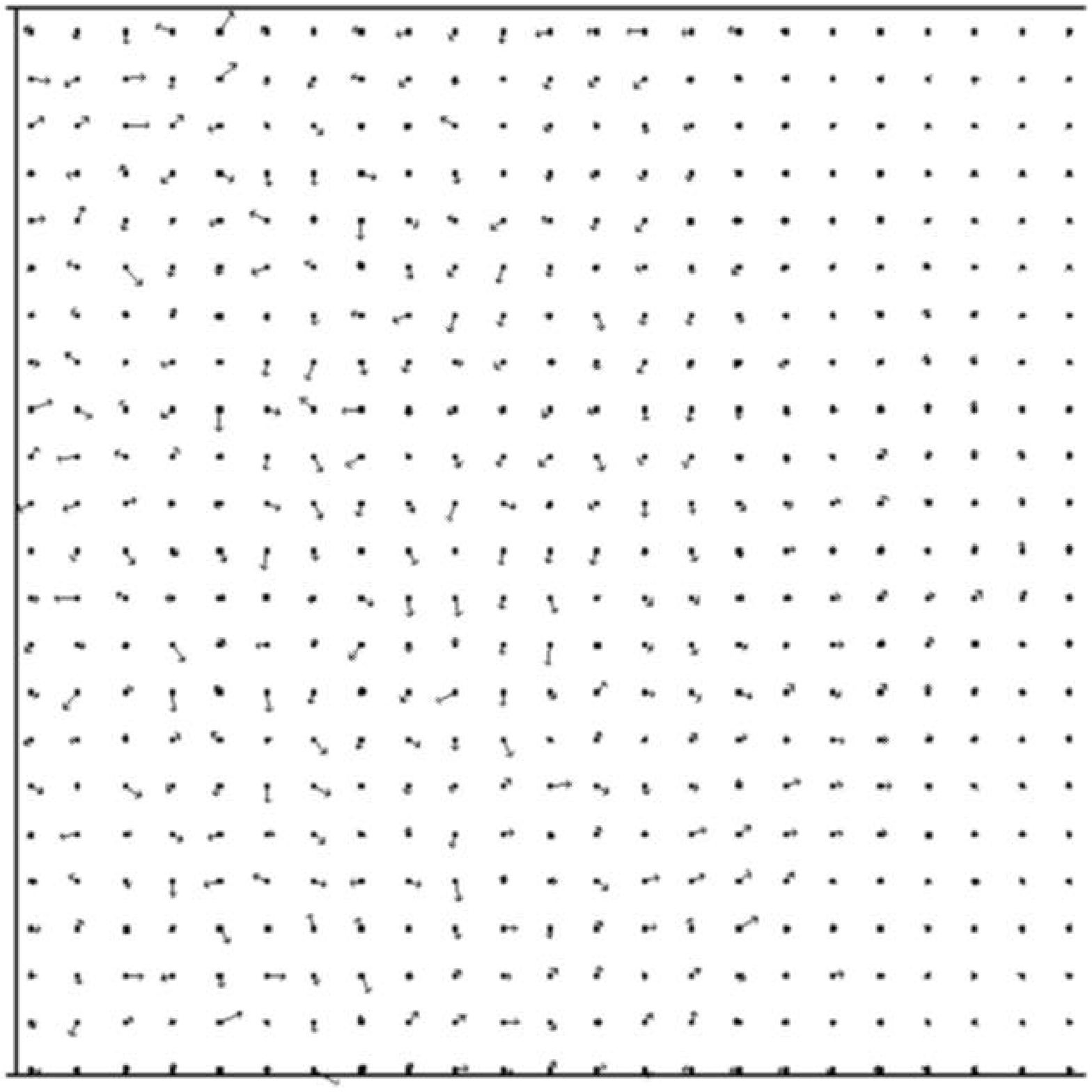}}
\caption{
Averaged velocity field (between $t_1=250$ and $t_2=300$) in a mixture
with $L_x=L_y=113$, $g=1$, $r_{11}=0.96$, $r_{22}=0.9992$, $N=2200$,
corresponding to $K=0.06$, $F_r=0.36$. Left frame corresponds to the
point A in figure~\ref{fig_circulation} (a doping value $\chi=0.11$),
while right frame refers to point B ($\chi=0.12$) }
\label{fig_impurity}
\end{figure}

In figure~\ref{fig_impurity} we show two different velocity fields
observed below and above the critical value of the doping. It is quite
striking to observe how the system behavior changes qualitatively in
correspondence of a tiny variation of the doping parameter. The simple
picture of an effective restitution coefficient (which in this case
should be a weigthed average between the two coefficients, using the
relative compositions as weights) could be an interpretation of the
results, as in the previous section. The transition observed in
figure~\ref{fig_circulation} could be explained as a transition driven
by the effective restitution coefficient when the weights are
changed. However the validation of such an hypothesis would require
systematic measurements, beyond the aim of the present investigation.

To conclude, we have explored the possibility of an hysteretic behavior:
the system has been prepared with a doping value $\chi$ slightly higher
than $\chi_{cr}$, but with an initial velocity field corresponding to
the convective cell at $\chi<\chi_{cr}$. We have studied the temporal
evolution of the circulation $\Phi(t)$: the circulation has initially a
finite value, but quickly decays to zero, indicating that the convection
is rapidly washed out. This behavior excludes therefore the possibility
of hysteresis or long transients.

\section{Conclusions}

In this paper, we have studied a two-dimensional granular gas by means
of Event-Driven numerical simulation. We employed a realistic model
regarding both the description of the grains as inelastic hard disks and
their interactions with the lateral walls and the horizontally
oscillating base.  The present results are in good qualitative agreement
with those predicted by the hydrodynamic
theory~\cite{meerson1,meerson2}. In particular, we observe the same
predicted dynamical regimes and confirm the non-trivial shape of the
phase diagram of~\cite{meerson2}, where the effects of inelasticity,
number of monolayers and gravity together influence the appearance or
disappearance of convection rolls.  Within the convective regime, the
coupling between the various fields (velocity field, density and
granular temperature) have been characterized.  Besides, we have
considered the influence of the bidispersity in composition by studying
mixtures with different types of grains. This is important in industrial
applications, where monodispersity is always difficult to be
obtained. We found that the addition of a small fraction of
quasi-elastic component is able to determine the disappearance of the
convective rolls. Our results suggest that a granular binary mixture of
grains with different inelasticities behaves, from the point of view of
convection, as a monodisperse system with an effective restitution
coefficient which is roughly the average (weighted with the composition
fractions of the two species) of the coefficients of the
components. Further investigations are necessary to make this statement
more precise.

\section{Acknowledgements}
The numerical code used in the present work was developed by Dr. Ciro
Cattuto. We wish to thank him for supporting the simulation work.

\end{document}